\documentclass[10pt,twoside,BCOR7mm,DIV15,headinclude,footexclude,
               cleardoubleempty,idxtotoc]
{scrartcl}

\usepackage{natbib}
\usepackage[font=small,labelfont=bf]{caption}
\usepackage[english]{babel}
\usepackage{graphicx}
\usepackage{hyperref}
\usepackage{scrpage2}
\usepackage{ifthen}
\usepackage{booktabs}
\usepackage{amsmath}
\usepackage{amssymb}
\usepackage{multicol}
\usepackage{float}
\usepackage{hyperref}

\hypersetup{breaklinks=true
,colorlinks=true,linkcolor=black,urlcolor=blue
,citecolor=black}

\addto\captionsenglish{%
}

\makeatletter
\renewcommand{\@biblabel}[1]{}
\renewcommand{\@cite}[2]{%
{#1\ifthenelse{\boolean{@tempswa}}{,#2}{}}}
\makeatother
\ofoot{\thepage}
\ifoot{}

\setcapindent{0em}
\setheadsepline{1pt}

\setkomafont{pagehead}{\normalfont\sffamily}
\setkomafont{pagenumber}{\normalfont\rmfamily}

\makeatletter
\newcommand{\listofcontributions}{\@starttoc{con}}

\newcommand{\l@contribution} {\@dottedtocline{1}{1.5em}{2.3em}}
\makeatother

\newenvironment{contribution}{
\setcounter{section}{0}
\setcounter{figure}{0}
\setcounter{table}{0}
}{
\newpage
\lehead{}
\rohead{}
}

\begin{document}

\setlength{\baselineskip}{2.5ex}

\begin{contribution}

\lehead{J.~L.\ Hoffman \& J.~R.\ Lomax}

\rohead{Structure of binary WR stars via spectropolarimetry}

\begin{center}
{\LARGE \bf Structure and fate of binary WR stars: \\ Clues from spectropolarimetry}\\
\medskip

{\it\bf Jennifer L. Hoffman$^1$ \& Jamie R.\ Lomax$^2$} \\

{\it $^1$Dept. of Physics \& Astronomy, University of Denver, USA}\\
{\it $^2$Homer L. Dodge Dept. of Physics \& Astronomy, University of Oklahoma, USA}

\begin{abstract}
Because most massive stars have been or will be affected by a companion during the course of their evolution, we cannot afford to neglect binaries when discussing the progenitors of supernovae and GRBs. Analyzing linear polarization in the emission lines of close binary systems allows us to probe the structures of these systems' winds and mass flows, making it possible to map the complex morphologies of the mass loss and mass transfer structures that shape their subsequent evolution. In Wolf-Rayet (WR) binaries, line polarization variations with orbital phase distinguish polarimetric signatures arising from lines that scatter near the stars from those that scatter far from the orbital plane. These far-scattering lines may form the basis for a ``binary line-effect method" of identifying rapidly rotating WR stars (and hence GRB progenitor candidates) in binary systems.
\end{abstract}
\end{center}

\begin{multicols}{2}

\section{Introduction}
In recent years, it has become clear that most massive stars evolve in close binary systems \citep[at least 75\%;][]{kiminki12updated,sana12binary,smith14mass}. Meanwhile, a growing number of core-collapse supernovae (SNe) show evidence for binary-induced asphericities in their ejecta and surroundings \citep[e.g., ][]{maund09early,chornock11transitional,mauerhan14multi}. Similarly, the necessity of rapid rotation to the formation of gamma-ray bursts (GRBs) via stripped-envelope SNe strongly suggests a binary origin as well (\cite{demink13rotation}. In short, ``it is no longer true that single WR stars are the preferred progenitors of most stripped-envelope SNe'' \citep{smith14mass}.

No progenitor (single or binary) of a stripped-envelope SN has yet been detected in pre-explosion images \citep{smartt09progenitors}. Many theoretical and statistical studies have considered binary channels for SN/GRB production \citep[e.g.,] []{podsiadlowski92presupernova,lyman14bolometric}, but observational corroboration relies primarily on nondetections and upper limits to progenitor masses  \citep[e.g., ][]{eldridge13death,kuncarayakti15nebular}. By contrast, the ``line-effect method" devised by \citet{vink07constraining} \textit{directly} identifies rapidly rotating Wolf-Rayet stars, which are the leading candidates for GRB progenitors under the collapsar model \citep{woosley93gamma,woosley13models}. 15 to 20\% of single Milky Way WR stars show a line effect \citep{vink07constraining,vink11pursuit,grafener12rotating}. However, because polarization arising from scattering in intra-binary circumstellar material (CSM) complicates the diagnostic potential of the line-effect method, this method has not been applied to WR binary systems. Given that most massive stars occur in binaries and that binary interactions dominate massive stellar evolution, the current statistics likely do not represent the full GRB progenitor population. 

\section{Polarization and the (Binary) Line-Effect Method}

Polarimetry provides direct information about an unresolved object's geometrical characteristics. Light from an unresolved spherical electron-scattering envelope is unpolarized due to cancellation of the electric vectors; however, in an aspherical envelope, incomplete cancellation produces a net linear polarization. Nonzero continuum polarization thus implies that the scattering region possesses a global asphericity. 

Polarization in spectral lines probes more complex scenarios. Line polarization signatures contain information about the geometries of individual elements in a stellar wind. In a rapidly rotating WR star, continuum photons form near the stellar surface and become polarized in the dense aspherical inner wind, while line photons form farther out in the wind and escape with little net polarization. This causes a depolarization across strong emission lines known as the ``line effect" \citep{vink07constraining}.  The asphericity revealed by this effect implies that the star meets the conditions necessary for GRB formation in the collapsar model.

Despite its diagnostic power, the line effect can be complicated by the effects of binarity. Mass exchange between stars in a binary system gives rise to complex CSM distributions that scatter starlight, producing time-variable continuum polarization signatures. Line photons may also scatter in this CSM and acquire intrinsic line polarization that confuses the line-effect diagnostic. For these reasons, studies of the line effect in WR stars have so far excluded binaries. However, because CSM scattering effects tend to depend on orbital phase, long-term monitoring can disentangle their contributions to WR emission-line polarization, allowing us to add WR binaries to the still-small sample of WR stars examined for the line effect.
\vfill

\begin{figure}[H]
\includegraphics[width=0.9\columnwidth]{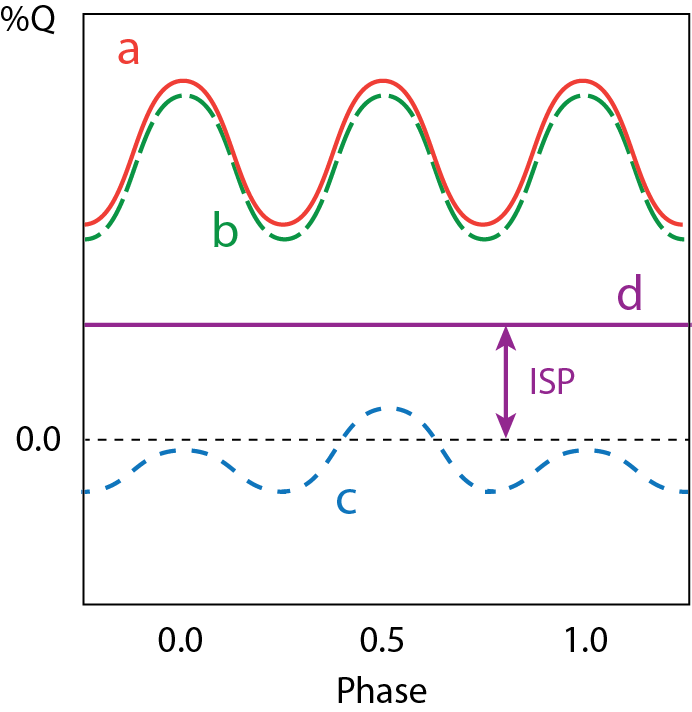}
\caption{Schematic polarimetric variation with binary phase of \textit{a)} the continuum; \textit{b)} an emission line arising and scattering like the continuum; \textit{c)} an example emission line arising and scattering differently from the continuum; and \textit{d)} an emission line arising and scattering far from the orbital plane. The dotted horizontal line marks the zero of Stokes \textit{Q}.}
\end{figure}

Figure 1 depicts the expected phase-dependent polarization behavior of different possible emission lines in a WR binary system. We show here the polarization of the line alone, without continuum contributions. We also display the $\%Q$ Stokes parameter only, taking a coordinate system aligned with the binary orbit. The continuum polarization (\textit{a}) varies sinusoidally with phase \citep{brown78polarisation,stlouis93polarization}. Polarization in an emission line whose photons formed near the stars and scattered near the orbital plane in the same way as the continuum photons (an unlikely scenario) would mirror that of the continuum (\textit{b}). More commonly, an emission line will both form and scatter differently from the continuum. If the scattering occurs near the orbital plane, it will still show a phase-locked variation, but the shape and magnitude of the polarization curve will be different from that of the continuum \citep[\textit{c}, for example, as in the He II $\lambda$5876 line of $\beta$ Lyr; ][]{lomax12geometrical}. Finally, an intrinsically unpolarized line forming and scattering far from the orbital plane should show a constant behavior with orbital phase (\textit{d}); its offset from zero polarization then provides a measure of the interstellar polarization (ISP). Such far-scattering lines have the potential to serve as line-effect diagnostics in binary WR systems. Of course, any given line may contain contributions from multiple, differently scattered components, but because of the vector nature of polarization, good phase coverage should allow the constant and variable components to be separated.

\begin{figure}[H]
\begin{center}
\includegraphics[width=\columnwidth]{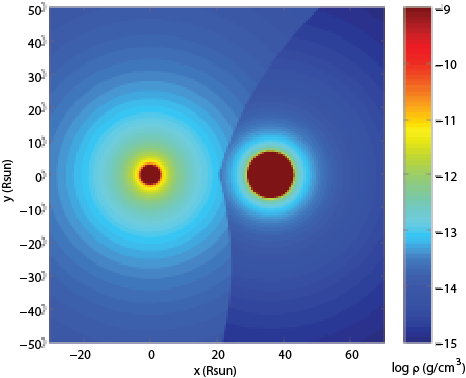}
\caption{Density in the orbital plane produced by the two-wind plus shock-cone model fitting the X-ray light curve of V444 Cyg \citep{lomax15v444}. The WN star is on the left and the O star is on the right. The wide opening angle of the shock cone is evidence for radiative braking/inhibition in the system.}
\end{center}
\end{figure}
\section{Test Case: V444 Cygni \\ (WR 139)}

V444 Cygni is a well studied, bright, eclipsing, close WN5+O6 binary system whose  colliding winds produce bright and variable X-ray emission. \citet{lomax15v444} reproduced the major features of the X-ray light curves with a model including weaker absorption at phases 0.3--0.75, when the cavity carved out by the O-star wind in the denser WR wind opens along the line of sight (Figure 2). This large phase range suggests that the shock cone has a wide opening angle, providing direct evidence for radiative braking and radiative inhibition effects within the system.

We observed V444 Cyg with the HPOL spectropolarimeter at Pine Bluff and Ritter Observatories \citep{davidson14hpol} and constructed line polarization phase curves for several emission lines \citep{lomax15v444}. Figure 3 shows the phase-dependent polarization behavior of these lines, each with continuum removed and rotated to its own average position angle. All lines show similar phase variations that are unlike those of the continuum \citep{stlouis93polarization}, implying that all contain intrinsic line polarization due to intra-binary scattering. All lines show a distinct difference in polarization between the phases ``in" and ``out" of the shock cone defined by X-ray modeling \citep{lomax15v444}. None of the lines appears to form far from the stars, though further work is needed to determine whether constant components exist. Additional analysis and modeling of these data will provide constraints on the stratified wind structure in V444 Cyg. 

\begin{figure*}[!t]
\begin{center}
\includegraphics
  [width=0.75\textwidth]{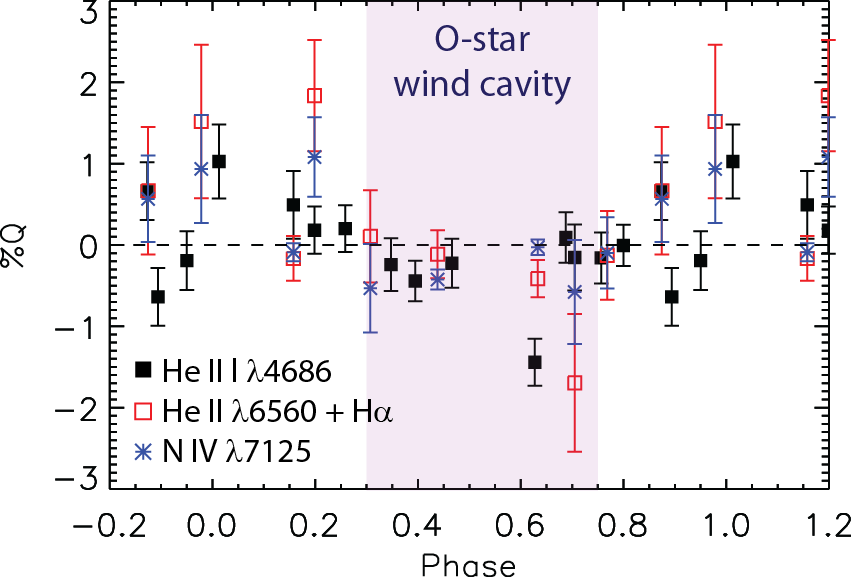}
\caption{Intrinsic line polarization for V444 Cyg, measured with HPOL. Data for each line have been rotated to their average PA to align the zero points of polarization. The shaded region denotes the interior of the shock cone produced by the colliding winds in the system, determined from modeling of the X-ray light curve \citep{lomax15v444}. Comparison with Fig. 1 suggests these lines all possess intrinsic polarization.
\label{example:bigfig}}
\end{center}
\end{figure*}

Continued spectropolarimetric monitoring of this and other WR binaries, supported by radiative transfer modeling (currently underway in Hoffman's group), will further test the utility of the binary line-effect method. We thank M. Corcoran, J. Davidson, M. de Becker, Y. Naz\'e, H. Neilson, S. Owocki, J. Pittard, A. Pollock, C. Russell, and the HPOL team for their contributions. This work has been supported by NSF award AST-1210372 and NASA ADAP award NNH12ZDA001N.

\bibliographystyle{aa} 
\bibliography{myarticle}

\end{multicols}
\end{contribution}


\end{document}